\documentclass{article}
\usepackage{amsmath}
\usepackage{graphicx}
\usepackage{epsfig}
\usepackage{amsfonts}
\usepackage{amssymb}
\usepackage{mathtools}

\newtheorem{theorem}{Theorem}

\newtheorem{proposition}[theorem]{Proposition}

\begin{document}

\title{Time-dependent cosmological interpretation of quantum mechanics}
\author{Emmanuel Moulay\thanks{XLIM (UMR-CNRS 7252), Universit\'{e} de Poitiers, 11 bd Marie et Pierre Curie, 86962 Futuroscope Chasseneuil Cedex, France; E-mail: emmanuel.moulay@univ-poitiers.fr}}
\date{}
\maketitle

\begin{abstract}
The aim of this article is to define a time-dependent cosmological interpretation of quantum mechanics in the context of an infinite open FLRW universe. A time-dependent quantum state is defined for observers in similar observable universes by using the particle horizon. Then, we prove that the wave function collapse of this quantum state is avoided.
\end{abstract}

\section{Introduction}

A new interpretation of quantum mechanics, called the cosmological interpretation of quantum mechanics, has been developed in order to take into account the new paradigm of eternal inflation \cite{Aguirre11,Moulay14,Vilenkin14}. Eternal inflation can lead to a collection of infinite open Friedmann-Lema\^{\i}tre-Robertson-Walker (FLRW) bubble universes belonging to a multiverse \cite{Coleman80,Garriga07,Guth07}. This inflationary scenario is called open inflation \cite{Linde98,Yamauchi11}. Such a multiverse implies that there exist an infinite number of observers belonging to similar observable universes which are indistinguishable inside an infinite open FLRW bubble universe \cite{Aguirre11,Tegmark03}. A quantum state $\left|\Psi_i\right\rangle\in\mathcal{H}$ is associated with each observer $i$ belonging to these similar observable universes and it is possible to define a quantum state gathering all these observers
\begin{equation}\label{psi}
\left|\Psi_\infty\right\rangle = \bigotimes_{i=1}^{\infty}  \left|\Psi_i\right\rangle
\end{equation}
which belongs to the Hilbert space
\begin{equation}
\mathcal{H}^{\otimes\infty}\coloneqq\mathcal{H}\otimes\mathcal{H}\otimes\mathcal{H}\otimes\cdots
\end{equation}
For sake of simplicity, we only consider pure states in this article. The reader may refer to \cite{Bub88,Gutmann95,VonNeumann39,VonNeumann01} for more details about the notion of infinite quantum states. Such a modelling is compatible with the Born rule \cite{Aguirre11,Farhi89,Finkelstein63,Hartle68,Vilenkin14} and can avoid the problem of wave function collapse \cite{Moulay14}. However, the notion of time is not well defined because similar observable universes are not causally related.

The first goal of this article is to define a time-dependent quantum state for observers in similar observable universes. This problem can be solved by using the particle horizon and the Fischler-Susskind cosmological holographic principle \cite{Fischler98,Susskind05}. The holographic principle has also been used in \cite{Bousso12,Nomura11} to render the many-worlds interpretation of quantum mechanics compatible with eternal inflation.

The second goal of this article is to prove that the wave function collapse of the time-dependent quantum state for observers in similar observable universes is avoided.

The organization of the article is the following. In Section \ref{Sec Time}, a time-dependent quantum state is defined for observers belonging to similar observable universes of an infinite open FLRW universe by using the particle horizon. The problem of wave function collapse of this quantum state is addressed in Section \ref{Sec Collapse}.

\section{Time-dependent quantum states}\label{Sec Time}

Let us consider two similar observers belonging to two similar observable universes of an infinite open FLRW universe which are indistinguishable. Let us denote by  $\left|\Psi_1\right\rangle$ the quantum state of the observer $1$ and by $\left|\Psi_2\right\rangle$ the  quantum state of the observer $2$. We may wonder if it is possible to define a common notion of time for these two observers. We want to use the time elapsed since the creation of the particle horizon as the same reference time. We may wonder if the two observers belonging to two similar observable universes at a given time have a similar past since the creation of the particle horizon.

The FLRW metric is given by
\begin{equation}
ds^2=c^2 dt^2-a(t)^2\left(\frac{dr^2}{1-k r^2}+r^2 d\Omega^2\right).
\end{equation}
and a natural definition of a cosmological horizon for a FLRW universe is the particle horizon whose proper radius is defined at time $t$ by
\begin{equation}
R_P(t)=a(t)\int_{t_i}^t \frac{c}{a(s)}ds
\end{equation}
where $t_i$ denotes the post-inflationary epoch \cite[Section 2.7]{Coles02}. The particle horizon is the largest comoving spatial distance from which light could have reached an observer if it was emitted at time $t_i$ \cite{Roos04}. It represents the boundary between the observable and the unobservable regions of the universe for an observer.

A particle cannot be ejected out of the observable universe of an observer by crossing his particle horizon \cite[page 37]{Lachieze99}. It implies that the existence of the particle horizon ensures that an observer has access to the information concerning the past of his observable universe since the creation of the particle horizon, even if there is a repulsive cosmological constant. The Fischler-Susskind cosmological holographic principle, which ensures that the particle horizon is compatible with special relativity \cite{Fischler98,Bak00}, states that the entropy of matter inside the post-inflationary particle horizon must be smaller than the area of the cosmological horizon \cite{Fischler98,Kaloper99}. It is true for open and classical flat FLRW universes \cite{Bousso02}. As open inflation leads to infinite open FLRW bubble universes \cite{Coleman80,Garriga07,Guth07}, the Fischler-Susskind cosmological holographic principle can be applied in this framework.

Thus, we have the following result:

\begin{proposition}
If two similar observers of an infinite open FLRW universe have indistinguishable observable universes at time $t_f$ after the post-inflationary epoch $t_i$ then they have similar observable universes since the post-inflationary epoch, i.e. the quantum states $\left|\Psi_1(t)\right\rangle$ and $\left|\Psi_2(t)\right\rangle$ are indistinguishable for all $t_i\leq t\leq t_f$.
\end{proposition}

We define the quantum state
\begin{equation}
\left|\Psi_{12}(t)\right\rangle=\left|\Psi_1(t)\right\rangle\otimes\left|\Psi_2(t)\right\rangle, \qquad t_i\leq t\leq t_f
\end{equation}
and we know that $\left|\Psi_1(t)\right\rangle$ and $\left|\Psi_2(t)\right\rangle$ are indistinguishable for all $t_i\leq t\leq t_f$. If we consider all the quantum states of all the observers having an observable universe similar to the observable universes 1 and 2 in an infinite open FLRW universe, then the generalization to the quantum state \eqref{psi} is straightforward and we obtain the time-dependent quantum state
\begin{equation}\label{psi time}
\left|\Psi_\infty(t)\right\rangle = \bigotimes_{i=1}^{\infty}  \left|\Psi_i(t)\right\rangle, \qquad t_i\leq t\leq t_f
\end{equation}
associated with the cosmological interpretation of quantum mechanics

Let us remark that it is possible to extend the existence of the particle horizon of an open FLRW universe to the Planck epoch by using string cosmology \cite{Bak00}.

\section{Wave function collapse}\label{Sec Collapse}

Wave function collapse associated with the quantum state \eqref{psi time} must be avoided after the post-inflationary epoch $t_i$. If there exists a meta-observer who is able to know the global result of a measurement process occurring in each similar observable universe of the multiverse at a given time then the measurement problem cannot be avoided. We have shown in \cite{Moulay14} that the collapse of the time-independent quantum state \eqref{psi} can be avoided. However, the same reasoning cannot be used for the time-dependent quantum state \eqref{psi time}. Indeed, if a measurement is done at time $t_m > t_i$, all quantum states $\left|\Psi_i(t)\right\rangle$ collapse at the same time $t_m$ after the post-inflationary epoch which was not the case for time-independent quantum states.

First, let us remark that if we have only a finite fixed number of similar observable universes, the wave function collapse cannot be avoided. Suppose that we have only a finite fixed number $N$ of similar observers $i$ in similar observable universes having quantum states $\left|\Psi_i(t)\right\rangle$. Just after the measurement at time $t^+_m=t_m+\epsilon$, we may have for instance
\begin{equation}
\left|\Psi_i(t^+_m)\right\rangle=\left|\Psi_j(t^+_m)\right\rangle
\end{equation}
for all $i,j\in\left\{1,\cdots,N\right\}$ where $\epsilon>0$ is a sufficiently small number. So, the quantum state $\bigotimes_{i=1}^{N}  \left|\Psi_i(t)\right\rangle$ collapses at time $t_m$ and all its possible evolutions cannot be explored. Let us remark that the Born rule is also not satisfied in a large but finite universe \cite{Page09,Page10} whereas it can be recovered in an infinite universe by using the frequency operator \cite{Aguirre11}.

Then, we prove that the wave function collapse of the time-dependent quantum state $\left|\Psi_\infty(t)\right\rangle$ is not possible. Let us denote by $t^-_m$ the time just before the measurement in each similar observable universe and $t^+_m$ the time just after the measurement. We study the quantum state $\left|\Psi_\infty(t^-_m)\right\rangle$ in order to see if its collapse is possible. As the number of quantum states in a finite region of the universe is finite \cite{Tegmark03}, there exists $1<K<+\infty$ such that for all $i\in\mathbb{N}^*=\mathbb{N}\setminus\{0\}$
\begin{equation}
\left|\Psi_i(t^-_m)\right\rangle =\sum_{k=1}^{K}\alpha_{ik}\left|\Psi_{k}(t^+_m)\right\rangle
\end{equation}
where $\alpha_{ik}\in\mathbb{C}$ and
\begin{equation}\label{Unitary}
\sum_{k=1}^{K}\left|\alpha_{ik}\right|^2=1.
\end{equation}
We have
\begin{equation}
\left|\Psi_\infty(t^-_m)\right\rangle =\bigotimes_{i=1}^{\infty}  \left|\Psi_i(t^-_m)\right\rangle=\lim_{N\rightarrow+\infty}\left(\bigotimes_{i=1}^{N} \sum_{k=1}^{K}\alpha_{ik}\left|\Psi_{k}(t^+_m)\right\rangle\right)
\end{equation}
We develop the previous expression in square brackets and we obtain
\begin{equation}
\left|\Psi_\infty(t^-_m)\right\rangle = \lim_{N\rightarrow+\infty}\left(\sum_{f_N\in \mathcal{T}_N} \bigotimes_{i=1}^{N}\alpha_{i f_N(i)}\left|\Psi_{f_N(i)}(t^+_m)\right\rangle\right)
\end{equation}
where $\mathcal{T}_N$ is the set of all the functions between $\left\{1,\cdots,N\right\}$ and $\left\{1,\cdots,K\right\}$. We also have $card(\mathcal{T}_N)=K^N$. Then, we gather the coefficients $\alpha$ by using the properties of the tensor product and it leads to
\begin{equation}
\left|\Psi_\infty(t^-_m)\right\rangle = \lim_{N\rightarrow+\infty}\left(\sum_{f_N\in \mathcal{T}_N}p^{f_N}_{N} \bigotimes_{i=1}^{N}\left|\Psi_{f_N(i)}(t^+_m)\right\rangle\right)
\end{equation}
with
\begin{equation}
p^{f_N}_{N}=\prod_{j=1}^{N}\alpha_{j f_N(j)}.
\end{equation}
The term
\begin{equation}
\left|p^{f_N}_{N}\right|^2=\left|\prod_{j=1}^{N}\alpha_{j f_N(j)}\right|^2=\prod_{j=1}^{N}\left|\alpha_{j f_N(j)}\right|^2
\end{equation}
is the probability of having the quantum state $\bigotimes_{i=1}^{N}\left|\Psi_{f_N(i)}(t^+_m)\right\rangle$ for the first $N$ observable universes. Let $\mathcal{T}$ be the set of all the functions between $\mathbb{N}^*$ and $\left\{1,\cdots,K\right\}$ and
\begin{equation}
p^f=\lim_{N\rightarrow+\infty}p^{f_N}_{N}=\prod_{j=1}^{\infty}\alpha_{j f(j)}
\end{equation}
with $f=\lim_{N\rightarrow+\infty}f_N\in\mathcal{T}$ then we have
\begin{equation}
\left|\Psi_\infty(t^-_m)\right\rangle = \sum_{f\in \mathcal{T}}p^f \bigotimes_{i=1}^{\infty}\left|\Psi_{f(i)}(t^+_m)\right\rangle.
\end{equation}
In \cite{Moulay14}, we have proved that
\begin{equation}\label{Eq}
\left|\alpha_{ik}\right|=\left|\alpha_{jk}\right|
\end{equation}
for all $i,j\in\mathbb{N}^*$ by using the Finkelstein-Hartle theorem \cite{Finkelstein63,Hartle68}. As
\begin{equation}\label{Ineq}
\left|\alpha_{ik}\right|^2<1
\end{equation}
for all $i\in\mathbb{N}^*$, $k\in\left\{1,\cdots,K\right\}$, we have
\begin{equation}\label{Collapse}
\lim_{N\rightarrow+\infty}\left|p^{f_N}_{N}\right|^2=\prod_{j=1}^{\infty}\left|\alpha_{j f(j)}\right|^2=0
\end{equation}
for all $f\in \mathcal{T}$. Indeed, a necessary condition for the product $\prod_{j=1}^{\infty}\left|\alpha_{j f(j)}\right|^2$ to be equal to a finite non zero positive real number is that
\begin{equation}\label{Condition}
\lim_{j\rightarrow+\infty}\left|\alpha_{j f(j)}\right|^2=1.
\end{equation}
The reader may refer to \cite[Chapter 2]{Melnikov11} for more details on infinite products. It is obvious that Condition \eqref{Condition} cannot be satisfied if we have \eqref{Eq} and \eqref{Ineq}. Let us remark that even if $K=+\infty$ which corresponds to an infinite number of quantum states in a finite volume, a necessary condition for \eqref{Unitary} is that $\lim_{k\rightarrow+\infty}\left|\alpha_{ik}\right|^2=0$ which is also not compatible with \eqref{Condition}.

Finally, the result given by \eqref{Collapse} leads to the following proposition:

\begin{proposition}
The probability of measuring the following quantum state
\begin{equation}
\bigotimes_{i=1}^{\infty}\left|\Psi_{f(i)}(t^+_m)\right\rangle
\end{equation}
is zero for all $f\in\mathcal{T}$. Thus, the collapse of the quantum state $\left|\Psi_\infty(t^-_m)\right\rangle$ is not possible.
\end{proposition}

We have defined a time-dependent quantum state $\left|\Psi_\infty(t)\right\rangle$ for observers in similar observable universes in Section \ref{Sec Time} and we may wonder why this quantum state does not collapse. This is due to the fact that the wave function collapse is associated with the notion of observer. In an infinite universe, this notion of observer falls down and then also the notion of wave function collapse. An observer $i$ can only see the wave function collapse of his quantum state $\left|\Psi_i(t)\right\rangle$ and he also knows that all the other observers in similar observable universes can see the wave function collapse of their quantum state at the same time $t_m$ elapsed since the post-inflationary epoch $t_i$ with a probability in accordance with the Born rule \cite{Aguirre11}. However, the global picture is not the wave function collapse of the quantum state $\left|\Psi_\infty(t)\right\rangle$ at time $t_m$ with the existence of a meta-observer.

\section*{Acknowledgements}
The author wants to thank Anthony Aguirre from the University of California for helpful discussions.

\bibliographystyle{elsarticle-num}
\bibliography{Time}

\end{document}